\documentclass[aps,prl,preprintnumbers,amsmath,amssymb,twocolumn,
superscriptaddress,showpacs,floatfix]{revtex4}
\usepackage{bbm}
\usepackage{mathrsfs}
\usepackage{amsmath,amssymb,bm,comment}
\usepackage{hyperref}
\usepackage{graphicx}
\usepackage{dcolumn} 
\usepackage{bm}

\begin{document}

\begin{flushright}
{\mbox{\hspace{10cm}}} USTC-ICTS-12-14  
\end{flushright}

\title{Berry curvature and 4-dimensional monopole in relativistic chiral kinetic equation}

\author{Jiunn-Wei Chen}
\affiliation{Department of Physics, National Center for Theoretical Sciences, and Leung Center for
Cosmology and Particle Astrophysics, National Taiwan University, Taipei
10617, Taiwan}

\author{Shi Pu}
\affiliation{Department of Physics, National Center for Theoretical Sciences, and Leung Center for
Cosmology and Particle Astrophysics, National Taiwan University, Taipei
10617, Taiwan}
\affiliation{Interdisciplinary Center for Theoretical Study and
Department of Modern Physics, University of Science and Technology
of China, Hefei 230026, China}

\author{Qun Wang}
\affiliation{Interdisciplinary Center for Theoretical Study and
Department of Modern Physics, University of Science and Technology
of China, Hefei 230026, China}

\author{Xin-Nian Wang}
\affiliation{Key Laboratory of Quark and Lepton Physics (MOE) 
and Institute of Particle Physics, Central China Normal University, Wuhan, 430079, China}
\affiliation{Nuclear Science Division, MS 70R0319, Lawrence Berkeley National Laboratory,
Berkeley, California 94720}

\date{\today}

\begin{abstract}
We derive a relativistic chiral kinetic equation with manifest Lorentz covariance 
from Wigner functions of spin-1/2 massless fermions in a constant background electromagnetic field.  
It contains vorticity terms and a  4-dimensional Euclidean Berry monopole which gives axial anomaly. By integrating out the zero-th component of the 4-momentum $p$, 
we reproduce the previous 3-dimensional results derived from the Hamiltonian approach, 
together with the newly derived vorticity terms. 
The phase space continuity equation has an anomalous source term 
proportional to the product of electric and magnetic fields 
($F_{\sigma\rho}\tilde{F}^{\sigma\rho} \sim  E_\sigma B^\sigma$).  
This provides a unified interpretation of the chiral magnetic
and vortical effects, chiral anomaly, Berry curvature, and the Berry monopole 
in the framework of Wigner functions. 
\end{abstract}


\maketitle

\textit{Introduction.} --- The Berry phase is a topological phase factor
acquired by an eigen-energy state when it undergoes adiabatic evolution
along a loop in parameter space \cite{Berry:1984}. It is in close analogy to
the Aharonov-Bohm phase when a charged particle moves in a loop enclosing a
magnetic flux, while the Berry curvature is like the magnetic field. The
integral of the Berry curvature over a closed surface can be quantized as
integers known as Chern-Simons numbers, which is similar to the Dirac
magnetic monopole and has deep connection with the quantum Hall effect. The
Berry phase is a beautiful, simple and universal structure in quantum
physics and has many interesting applications, for a recent review of the
Berry phase in condensed matter physics, see e.g. Ref. \cite{Xiao:2010}.

Recently it has been found that features of the Berry phase due to a
3-dimensional momentum monopole emerge in a chiral kinetic equation without
manifest Lorentz covariance \cite{Son:2012wh,Stephanov:2012ki}. A
semi-classical kinetic equation has also been derived in an electron system
with Berry curvature \cite{Wong:2011}. 
Chiral anomaly is an important quantum effect which is absent at the
classical level. It is manifested in the chiral magnetic and vortical effect
(CME and CVE) \cite{Kharzeev:2007jp,Fukushima:2008xe,Kharzeev:2010gr} as
electric currents induced by magnetic field and vorticity. Such effects and
related topics have been investigated within a variety of approaches, such
as AdS/CFT correspondence \cite%
{Erdmenger:2008rm,Banerjee:2008th,Torabian:2009qk,Rebhan:2009vc,Kalaydzhyan:2011vx}%
, relativistic hydrodynamics \cite%
{Son:2009tf,Pu:2010as,Sadofyev:2010pr,Lin:2011mr,Kharzeev:2011ds}, and
quantum field theory \cite{Metlitski:2005pr,Newman:2005as,Fukushima:2008xe,
Charbonneau:2009ax,Lublinsky:2009wr,Asakawa:2010bu,Landsteiner:2011cp,Hou:2012xg}%
.

In this paper we will derive a new chiral kinetic equation with manifest
Lorentz covariance from the Wigner function \cite{Gao:2012ix}. Such an
equation can provide semi-classical description of quantum transport
phenomena. We will show that such a chiral kinetic equation incorporates
features of the Berry curvature and 4-dimensional Euclidean monopole. These
results reveal the inherent connection between the Berry phase and gauge
invariant Wigner functions. One advantage of our approach is that the
vorticity effect in the chiral kinetic equation can be derived
straightforwardly which, apparently, is not the case in other approaches. We
also show that the previous non-covariant kinetic equation 
\cite{Son:2012wh,Stephanov:2012ki} is equivalent to our covariant one in the case
of zero vorticity after we integrate over the zero-th momentum $p_0$. The
relativistic chiral kinetic equation in our approach is quite general and
valid not only for Fermi liquid as in Ref.\ \cite%
{Son:2012wh,Stephanov:2012ki} but for any relativistic fermionic system. The
phase space continuity or Liouville's theorem can be shown to be broken by
an anomalous term proportional to the product of electric and magnetic
fields. So the phase space measure is not conserved. It is modified by a
factor related to the Berry curvature. We will also show that the
conservation law of the right- and left-hand currents is broken by anomalous
terms, which can be given by the flux of a 4-dimensional monopole in
Euclidean momentum space. Therefore we provide a unified interpretation of a
variety of properties such as CME/CVE, chiral anomaly, Berry curvature and
4-d Euclidean monopole in the framework of Wigner functions. We will use the
metric convention $g^{\mu\nu}=\mathrm{diag}(1,-1,-1,-1)$.

\textit{Equation of motion with Berry curvature in 3-dimension. } --- We
will follow an example in Ref.\ \cite{Son:2012wh} to illustrate the concept
of Berry curvature. We consider a Hamiltonian $H^{\prime }=\bm{\sigma}\cdot 
\mathbf{p}$ for spin-1/2 fermions in addition to the normal part $H(\mathbf{p%
},\mathbf{x})$, where $\bm{\sigma}$ are Pauli matrices. Under an adiabatic
evolution, the path-integral action for fermions with positive helicity is 
\begin{equation}
S=\int dt(\dot{\mathbf{x}}\cdot \mathbf{p}+\dot{\mathbf{x}}\cdot \mathbf{A}(%
\mathbf{x})-\dot{\mathbf{p}}\cdot \mathbf{a}(\mathbf{p})-H(\mathbf{p},%
\mathbf{x})],
\end{equation}%
where $\mathbf{A}(\mathbf{x})$ is the electromagnetic vector potential and $%
\mathbf{a}(\mathbf{p})$ is the vector potential in momentum space resulting
from diagonalizing $H^{\prime }$ in path integral. We can generalize the
coordinate variables by combining $\mathbf{p}$ and $\mathbf{x}$, $\xi _{a}=(%
\mathbf{p},\mathbf{x})$ with $a=1,2,\cdots ,6$. The action can be cast into
a compact form, 
\begin{equation}
S=\int dt[-\gamma _{a}(\xi )\dot{\xi}_{a}-H(\xi )],
\end{equation}%
where $\gamma _{a}=[\mathbf{a}(\mathbf{p}),-\mathbf{p}-\mathbf{A}(\mathbf{x}%
)]$. The equations of motion are 
\begin{equation}
\gamma _{ab}\dot{\xi}_{b}=-\frac{\partial H(\xi )}{\partial \xi _{a}}
\end{equation}%
where $\gamma _{ab}\equiv \partial _{a}\gamma _{b}(\xi )-\partial _{b}\gamma
_{a}(\xi )$ is given by 
\begin{equation}
\lbrack \gamma _{ab}]=\left[ 
\begin{array}{cccccc}
0 & \Omega _{3} & -\Omega _{2} & -1 & 0 & 0 \\ 
-\Omega _{3} & 0 & \Omega _{1} & 0 & -1 & 0 \\ 
\Omega _{2} & -\Omega _{1} & 0 & 0 & 0 & -1 \\ 
1 & 0 & 0 & 0 & -B_{3} & B_{2} \\ 
0 & 1 & 0 & B_{3} & 0 & -B_{1} \\ 
0 & 0 & 1 & -B_{2} & B_{1} & 0%
\end{array}%
\right] ,
\end{equation}%
where $\bm{\Omega}=\nabla _{\mathbf{p}}\times \mathbf{a}(\mathbf{p})$ is the
Berry curvature and $\mathbf{B}=\nabla \times \mathbf{A}(\mathbf{x})$ is the
3-dimensional magnetic field. The determinant of $[\gamma _{ab}]$ is $\det
[\gamma _{ab}]=(1+\bm{\Omega}\cdot \mathbf{B})^{2}$. We see that the
invariant phase space volume becomes $\sqrt{\det [\gamma _{ab}]}d^{3}\mathbf{%
x}d^{3}\mathbf{p}$, where $\sqrt{\det [\gamma _{ab}]}=|1+\bm{\Omega}\cdot 
\mathbf{B}|$ indicates the change of phase space volume with time \cite%
{Xiao:2005qw}.

\textit{The Wigner function approach} --- In an alternative
quantum kinetic theory approach, the classical phase-space distribution 
$f(x,p)$ is replaced by the Wigner function $W(x,p)$ in space-time $x$ and
4-momentum $p$, defined as the ensemble average of the Wigner operator 
\cite{Elze:1986qd,Vasak:1987um,Elze:1989un} for spin-1/2 fermions, 
\begin{equation}
\hat{W}_{\alpha \beta }=\int \frac{d^{4}y}{(2\pi )^{4}}e^{-ip\cdot y}
\bar{\psi}_{\beta }(x_{+})U(x_{+},x_{-})\psi _{\alpha }(x_{-}),  
\label{wigner}
\end{equation}
where $\psi _{\alpha }$\ and $\bar{\psi}_{\beta }$ are Dirac spinor fields, 
$x_{\pm }\equiv x\pm \frac{1}{2}y$\ are two space-time points centered at 
$x$ with space-time separation $y$, and the gauge link $U$, 
\begin{equation}
U(x_{+},x_{-})\equiv e^{-iQ\int_{x_{-}}^{x_{+}}dz^{\mu }A_{\mu }(z)},
\label{link}
\end{equation}
ensures the gauge invariance of $\hat{W}_{\alpha \beta }$. Here $Q$\ is the
electromagnetic charge of the fermions, and $A_{\mu }$\ is the
electromagnetic vector potential. To simplify the quantum kinetic equation
under a background field we consider a massless and collisionless fermionic
system in a constant external electromagnetic field $F_{\mu \nu }$ in the
lab frame. The Wigner function for spin-1/2 fermions is a matrix in Dirac
space and satisfies the quantum kinetic equation 
\cite{Elze:1986qd,Vasak:1987um,Elze:1989un}, 
$\gamma _{\mu }(p^{\mu }+\frac{i}{2}\nabla ^{\mu })W(x,p)=0$, 
where $\gamma ^{\mu }$'s are Dirac matrices and 
$\nabla ^{\mu }\equiv \partial _{x}^{\mu }-Q{F^{\mu }}_{\nu }\partial
_{p}^{\nu }$. The Wigner function can be decomposed in terms of 16
independent generators of the Clifford algebra whose coefficients are
scalar, pseudo-scalar, vector, axial vector and tensor respectively. The
vector $\mathscr{V}_{\mu }(x,p)$ and axial-vector $\mathscr{A}_{\mu }(x,p)$
component of the Wigner function can be determined by the quantum kinetic
equations (i.e. Eqs.\ (5-8) of Ref.\ \cite{Gao:2012ix}) to the first order
of space-time derivative $\partial _{x}$ and the field strength $F_{\mu \nu }
$: 
\begin{eqnarray}
\mathscr{Z}^{\mu } &=&p^{\mu }\delta (p^{2})Z_{0}+\frac{1}{2}p_{\nu }[u^{\mu
}\omega ^{\nu }-u^{\nu }\omega ^{\mu }]\frac{\partial \bar{Z_{0}}}{\partial
(u\cdot p)}\delta (p^{2})  \notag  \label{V-final} \\
&&-Qp_{\nu }[u^{\mu }B^{\nu }-u^{\nu }B^{\mu }]\bar{Z}_{0}\delta ^{\prime
}(p^{2})  \notag \\
&&+Q\epsilon ^{\mu \lambda \rho \sigma }u_{\lambda }p_{\rho }E_{\sigma }\bar{%
Z}_{0}\delta ^{\prime }(p^{2}),
\end{eqnarray}%
where $\mathscr{Z}=(\mathscr{V},\mathscr{A})$, $Z_{0}=(V_{0},A_{0})$, $\bar{%
Z_{0}}=(A_{0},V_{0})$, with the first order solutions $V_{0}$ and $A_{0}$
given by 
\begin{eqnarray}
\left[ V_{0},A_{0}\right]  &=&\sum_{s=\pm 1}\theta (su\cdot p)\left[
(f_{s,R}+f_{s,L}),(f_{s,R}-f_{s,L})\right] ,  \notag  \label{disV} \\
f_{s,\chi } &=&\frac{2}{(2\pi )^{3}}\frac{1}{e^{s(u\cdot p-\mu _{\chi })/T}+1%
},(\chi =R,L),
\end{eqnarray}%
where $R(L)$ denotes the right (left)-handed fermions and $\mu _{R,L}=\mu
\pm \mu _{5}$. We have used notations $E_{\sigma }=u^{\rho }F_{\sigma \rho }$%
, $B_{\sigma }=(1/2)\epsilon _{\sigma \mu \nu \rho }u^{\mu }F^{\nu \rho }$
and $\omega _{\mu }=(1/2)\epsilon _{\mu \nu \rho \sigma }u^{\nu }\partial
^{\rho }u^{\sigma }$, which depend on $x$ only via the fluid velocity $u(x)$%
. We use $\mathscr{Z}_{0}^{\mu }$ to denote the zero-th order term $p^{\mu
}\delta (p^{2})Z_{0}$ in Eq.~(\ref{V-final}) and $\mathscr{Z}_{1}^{\mu }$
for the first order terms.

The vector and axial-vector current and the energy-momentum tensor can be
derived from $\mathscr{V}^\mu$ and $\mathscr{A}^\mu$ in Eq.~(\ref{V-final})
by integrating over momentum: $j^\mu =\int d^4p \mathscr{V}^\mu$, $j_5^\mu =
\int d^4p \mathscr{A}^\mu $, and $T^{\mu\nu} = \frac 12 \int d^4p (p^\mu %
\mathscr{V}^\nu + p^\nu \mathscr{V}^\mu)$. The current $j^\mu$ contains two
parts proportional to magnetic field and vorticity, known as the CME and CVE 
\cite{Kharzeev:2007jp,Fukushima:2008xe,Kharzeev:2010gr,Son:2009tf},
respectively. So both effects are contained in the Wigner function \cite%
{Gao:2012ix}. These currents and $T^{\mu\nu}$ are shown to obey conservation
equations \cite{Gao:2012ix}: $\partial_\mu j^\mu =0$, $\partial_\mu j_5^\mu
= - \frac{Q^2}{2\pi^2} E\cdot B$, and $\partial _\mu T^{\mu\nu}=QF^{\nu\rho}
j_\rho$.

\textit{Lorenz covariant chiral kinetic equation.} --- Now we try to derive
a new form of Lorentz covariant chiral kinetic equation in 4-dimensions from
Eq.\ (6) of Ref.\ \cite{Gao:2012ix}, i.e. $\nabla _{\mu }\mathscr{Z}^{\mu
}=0 $, which holds for the zero-th and first order Wigner function 
$\mathscr{Z}_{0}^{\mu }$ and $\mathscr{Z}_{1}^{\mu }$ separately. The zero-th
order equation $\nabla _{\mu }\mathscr{Z}_{0}^{\mu }=0$ can be rewritten
as, 
\begin{eqnarray}
\nabla _{\mu }\mathscr{Z}_{0}^{\mu } &=&(\partial _{\mu }^{x}-QF_{\mu \nu
}\partial _{p}^{\nu })[p^{\mu }\delta (p^{2})Z_{0}]  \notag \\
&=&\delta (p^{2})[p^{\mu }\partial _{\mu }^{x}-Qp^{\mu }F_{\mu \nu }\partial
_{p}^{\nu }]Z_{0}=0.  \label{eq:kin-v0}
\end{eqnarray}%
Here $Z_{0}$ is the phase space distribution function and given in Eq.~(\ref%
{disV}). Eq. (\ref{eq:kin-v0}) is a Vlasov-like equation, from which we can
extract $dx^{\sigma }/d\tau =p^{\sigma }/m_{0}$ and $dp^{\mu }/d\tau
=Qp^{\mu }F_{\mu \nu }/m_{0}$, where $\tau $ is a world-line parameter, $%
m_{0}$ is a quantity with mass dimension (but not the fermion mass since we
are considering massless fermions) which can be scaled away by collision
terms, and $Qp^{\mu }F_{\mu \nu }/m_{0}$ is a general Lorentz force.

Now we rewrite the first order equation $\nabla _{\mu }\mathscr{Z}_{1}^{\mu
}=0$ in a way that only terms of spatial and momentum derivatives of $\bar{Z}%
_{0}$ are kept, 
\begin{eqnarray}
\nabla _{\mu }\mathcal{Z}_{1}^{\mu } &=&Q\delta (p^{2})[(u\cdot b)B^{\mu
}-(b\cdot B)u^{\mu }  \notag \\
&&+\epsilon ^{\mu \nu \rho \sigma }u_{\nu }b_{\rho }E_{\sigma }]\partial
_{\mu }\bar{Z}_{0}  \notag \\
&&+\delta (p^{2})\left[ \frac{1}{2}\omega ^{\mu }+\omega ^{\mu }(p\cdot
u)(b\cdot u)\right.  \notag \\
&&\left. -2u^{\mu }(p\cdot \omega )(b\cdot u)\right] \partial _{\mu }\bar{Z}%
_{0}  \notag \\
&&-Q^{2}\delta (p^{2})(E\cdot B)b^{\sigma }\partial _{\sigma }^{p}\bar{Z}_{0}
\notag \\
&&+Q\delta (p^{2})\left[ \frac{1}{2}(\omega \cdot E)u^{\sigma }\right. 
\notag \\
&&+\left. \frac{1}{p^{2}}(p\cdot \omega )p_{\eta }F^{\sigma \eta }\right]
\partial _{\sigma }^{p}\bar{Z}_{0}=0\;.  \label{eq:kin-v1}
\end{eqnarray}%
where $b^{\sigma }\equiv -p^{\sigma }/p^{2}$. The first two terms are
proportional to $\partial _{\mu }\bar{Z}_{0}$ and the last two terms are
proportional to $\partial _{\sigma }^{p}\bar{Z}_{0}$. We will show that $%
\delta (p^{2})b^{\sigma }$ is a 4-dimensional monopole in Euclidean momentum
space. Combining Eq.~(\ref{eq:kin-v0}) and (\ref{eq:kin-v1}), we obtain the
Lorentz covariant chiral kinetic equation 
\begin{eqnarray}
&&\frac{1}{2}\nabla _{\mu }(\mathscr{V}^{\mu }\pm \mathscr{A}^{\mu })=0 
\notag \\
&\rightarrow &\delta (p^{2})\left[ \frac{dx^{\sigma }}{d\tau }\partial
_{\sigma }^{x}+\frac{dp^{\sigma }}{d\tau }\partial _{\sigma }^{p}\right]
f_{R/L}=0,  \label{eq:c-kin-eq}
\end{eqnarray}%
where the upper/lower sign corresponds to the right/left-hand distribution,
and $dx^{\sigma }/d\tau $ and $dp^{\sigma }/d\tau $ are given by 
\begin{eqnarray}
m_{0}\frac{dx^{\sigma }}{d\tau } &=&p^{\sigma }\pm Q\left[ (u\cdot
b)B^{\sigma }-(b\cdot B)u^{\sigma }+\epsilon ^{\sigma \alpha \beta \gamma
}u_{\alpha }b_{\beta }E_{\gamma }\right]  \notag \\
&&\pm \left[ \frac{1}{2}\omega ^{\sigma }+\omega ^{\sigma }(p\cdot u)(b\cdot
u)-2u^{\sigma }(p\cdot \omega )(b\cdot u)\right] ,  \notag \\
m_{0}\frac{dp^{\sigma }}{d\tau } &=&-Qp_{\rho }F^{\rho \sigma }\mp
Q^{2}(E\cdot B)b^{\sigma }  \notag \\
&&\pm Q\frac{1}{2}(\omega \cdot E)u^{\sigma }\mp Q(p\cdot \omega )b_{\eta
}F^{\sigma \eta }\;.  \label{dx-dp-dtau}
\end{eqnarray}%
Here we have used notations 
\begin{equation}
f_{R/L}\equiv \frac{1}{2}(V_{0}\pm A_{0})=\sum_{s=\pm 1}\theta (su\cdot
p)f_{s,R/L}.  \label{frl}
\end{equation}%
Using the spatial and momentum divergences, 
\begin{eqnarray}
\partial _{\sigma }\left[ \frac{dx^{\sigma }}{d\tau }\delta (p^{2})\right]
&=&0,  \notag \\
\partial _{\sigma }^{p}\left[ \frac{dp^{\sigma }}{d\tau }\delta (p^{2})%
\right] &=&\mp Q^{2}(E\cdot B)\partial _{\sigma }^{p}[b^{\sigma }\delta
(p^{2})],
\end{eqnarray}%
we obtain 
\begin{eqnarray}
&&\partial _{\sigma }\left[ \frac{dx^{\sigma }}{d\tau }\delta (p^{2})\right]
f_{R/L}+\partial _{\sigma }^{p}\left[ \frac{dp^{\sigma }}{d\tau }\delta
(p^{2})\right] f_{R/L}  \notag \\
&=&\mp Q^{2}(E\cdot B)\partial _{\sigma }^{p}[b^{\sigma }\delta
(p^{2})]f_{R/L}.  \label{div-1}
\end{eqnarray}%
It is interesting to see that $\delta (p^{2})dx^{\sigma }/d\tau $ is
conserved but $\delta (p^{2})dp^{\sigma }/d\tau $ is not. We can combine
Eq.\ (\ref{div-1}) with the chiral kinetic equation (\ref{eq:c-kin-eq}) to
obtain the Liouville equation or the phase space continuity equation, 
\begin{eqnarray}
&&\partial _{\sigma }\left[ \frac{dx^{\sigma }}{d\tau }\delta (p^{2})f_{R/L}%
\right] +\partial _{\sigma }^{p}\left[ \frac{dp^{\sigma }}{d\tau }\delta
(p^{2})f_{R/L}\right]  \notag \\
&=&\mp Q^{2}(E\cdot B)\partial _{\sigma }^{p}[b^{\sigma }\delta
(p^{2})]f_{R/L}.  \label{liouville}
\end{eqnarray}%
In deriving Eqs.\ (\ref{eq:c-kin-eq}-\ref{liouville}), we have used the
conditions $u^{\mu }\partial _{\mu }\omega ^{\nu }=u^{\mu }\partial _{\mu
}B^{\nu }=0$, $\partial _{\mu }\omega ^{\mu }=0$, $\partial _{\mu }B^{\mu
}=2(\omega \cdot E)$, $\partial ^{\mu }u^{\nu }=\epsilon ^{\mu \nu \tau
\lambda }u_{\tau }\omega _{\lambda }$, $\epsilon ^{\sigma \rho \eta \xi
}u_{\rho }\partial _{\sigma }E_{\xi }=0$, $\epsilon _{\mu \nu \sigma \rho
}u^{\mu }\omega ^{\nu }B^{\sigma }=0$, $\epsilon ^{\sigma \rho \alpha \beta
}\epsilon _{\sigma \rho \mu \nu }=-2\delta _{\lbrack \mu \nu ]}^{\alpha
\beta }$, $\partial _{\sigma }^{p}b^{\sigma }=-2/p^{2}$, and $\partial
_{\sigma }^{p}[\delta (p^{2})]=2b_{\sigma }\delta (p^{2})$. We see in Eq.\ (%
\ref{liouville}) the breaking of continuity for the phase space density by
an anomalous term proportional to $E\cdot B$.

The vector currents for right- and left-hand fermions can be given by
integration over 4-momentum from $dx^{\sigma}/d\tau$ as 
\begin{eqnarray}
j_{R/L}^{\sigma} & = & \int d^{4}p\delta(p^{2})\frac{dx^{\sigma}}{d\tau}%
f_{R/L}  \notag \\
& = & \frac{1}{2}(j^{\sigma}\pm j_{5}^{\sigma}),  \label{jrl}
\end{eqnarray}
where $j^{\sigma}$ and $j_{5}^{\sigma}$ are given in Eqs.\ (17-18) of Ref.\ 
\cite{Gao:2012ix}. The energy-momentum tensor can also be obtained from $%
dx^{\sigma}/d\tau$, 
\begin{equation}
T^{\sigma\rho} = \int d^4 p \left( p^\sigma \frac{dx^{\rho}}{d\tau} +p^\rho 
\frac{dx^{\sigma}}{d\tau} \right)
\end{equation}
which gives Eq.\ (19) of Ref.\ \cite{Gao:2012ix}. Note that the vorticity
terms in Eqs. (\ref{eq:c-kin-eq},\ref{dx-dp-dtau}) are necessary for the
presence of the CVE in $j^\sigma_{R/L}$ and $T^{\sigma\rho}$.

\textit{Chiral kinetic equation in 3-dimensions.} --- We can obtain the
chiral kinetic equation in 3-dimensions by integration over $p_{0}$ for the
Lorentz covariant chiral kinetic equation (\ref{eq:c-kin-eq}) as, 
\begin{equation}
\int dp_{0}\delta (p^{2})\left[ \frac{dx^{\sigma }}{d\tau }\partial _{\sigma
}^{x}f_{R/L}+\frac{dp^{\sigma }}{d\tau }\partial _{\sigma }^{p}f_{R/L}\right]
=0,  \label{int-cke-3}
\end{equation}%
which amounts to calculating the following integrals 
\begin{equation*}
I_{n}=\int dp_{0}\delta (p^{2})\frac{p_{0}^{n}}{p^{2}}F(x,p),
\end{equation*}%
with $n=0,1,2$. We use the $i\epsilon $ prescription 
\begin{equation}
\delta (x)\mathscr{P}\frac{1}{x}=-\frac{1}{2\pi }\mathrm{Im}\frac{1}{%
(x+i\epsilon )^{2}}
\end{equation}%
to evaluate the integrals by enclosing the pole $p_{0}=|\mathbf{p}%
|-i\epsilon $ in the lower half-plane of $p_{0}$, where $\epsilon $ is a
small positive number. Then we can derive the chiral kinetic equation in
3-dimensions from Eq.\ (\ref{int-cke-3}), 
\begin{equation}
\frac{dt}{d\tau }\partial _{t}f_{R/L}+\frac{d\mathbf{x}}{d\tau }\cdot \nabla
_{\mathbf{x}}f_{R/L}+\frac{d\mathbf{p}}{d\tau }\cdot \nabla _{\mathbf{p}%
}f_{R/L}=0,
\end{equation}%
where $dt/d\tau $, $d\mathbf{x}/d\tau $ and $d\mathbf{p}/d\tau $ are given
by 
\begin{eqnarray}
\frac{dt}{d\tau } &=&1\pm Q\bm{\Omega}\cdot \mathbf{B}\pm 4|\mathbf{p}|(%
\bm{\Omega}\cdot \bm{\omega}),  \notag \\
\frac{d\mathbf{x}}{d\tau } &=&\hat{\mathbf{p}}\pm Q(\hat{\mathbf{p}}\cdot %
\bm{\Omega})\mathbf{B}\pm Q(\mathbf{E}\times \bm{\Omega})\pm \frac{1}{|%
\mathbf{p}|}\bm{\omega},  \notag \\
\frac{d\mathbf{p}}{d\tau } &=&Q(\mathbf{E}+\hat{\mathbf{p}}\times \mathbf{B}%
)\pm Q^{2}(\mathbf{E}\cdot \mathbf{B})\bm{\Omega}  \notag \\
&&\mp Q|\mathbf{p}|(\mathbf{E}\cdot \bm{\omega})\bm{\Omega}\pm 3Q(\bm{\Omega}%
\cdot \bm{\omega})(\mathbf{p}\cdot \mathbf{E})\hat{\mathbf{p}},
\label{3-d-cke}
\end{eqnarray}%
with $\hat{\mathbf{p}}=\mathbf{p}/|\mathbf{p}|$ and the 3-dimensional Berry
curvature $\bm{\Omega}=\mathbf{p}/(2|\mathbf{p}|^{3})$ \cite%
{Son:2012wh,Stephanov:2012ki}. Turning off the $\omega $ terms in Eq.\ (\ref%
{3-d-cke}), we reproduce Eqs.~(14-15) in Ref.~\cite{Stephanov:2012ki}. Note
that the $\bm{\omega}$ terms in the chiral kinetic equation in 3-dimensions
come naturally from the Lorentz covariant chiral kinetic equation in
4-dimensions.

\textit{Anomaly and 4-dimensional Berry monopole.} --- From Eqs.~(\ref%
{V-final}), the anomalous conservation law of the left- and right-hand
current can be derived from $\partial _{\mu }j^{\mu }=0$ and $\partial _{\mu
}j_{5}^{\mu }=-\frac{Q^{2}}{2\pi ^{2}}E\cdot B$, 
\begin{equation}
\partial _{\rho }j_{R/L}^{\rho }=\mp \frac{Q^{2}}{4\pi ^{2}}(E\cdot B).
\label{anomaly-rl}
\end{equation}%
On the other hand, we can understand the chiral anomaly in Eq.~(\ref%
{anomaly-rl}) from the perspective of a 4-dimensional Berry monopole. To
this end, we act $\partial _{\sigma }$ on Eq.\ (\ref{jrl}), use Eqs.\ (\ref%
{liouville}) and carry out the integral in Euclidean space, 
\begin{eqnarray}
\partial _{\sigma }j_{R/L}^{\sigma } &=&\mp Q^{2}(E\cdot B)\int
d^{4}p\partial _{\sigma }^{p}[b^{\sigma }\delta (p^{2})]f_{R/L}  \notag \\
&=&\mp Q^{2}(E\cdot B)\frac{1}{\pi }\mathrm{Im}\int_{-i\infty }^{i\infty
}dp_{0}d^{3}p  \notag \\
&&\times \partial _{\sigma }^{p}\left[ \frac{p^{\sigma }}{p^{2}}\frac{1}{%
p^{2}+i\epsilon }\right] f_{R/L}  \notag \\
&=&\pm Q^{2}(E\cdot B)\frac{1}{\pi }\int_{-\infty }^{\infty
}dp_{4}d^{3}p\partial _{\sigma }^{p_{E}}\left[ \frac{p_{E}^{\sigma }}{%
p_{E}^{4}}\right] f_{R/L}  \notag \\
&=&\pm \frac{Q^{2}}{4\pi ^{2}}(E\cdot B).  
\label{euclidean}
\end{eqnarray}%
We have used $\pi \delta (x)=-\mathrm{Im}[1/(x+i\epsilon )]$ and taken
analytic continuation $p_{4}=ip_{0}$ and $p^{2}=-p_{E}^{2}$. Note that in
the second equality of Eq.~(\ref{euclidean}) the poles in Minkowski space
are $p_{0}=\pm \sqrt{|\mathbf{p}|^{2}-i\epsilon }=\pm |\mathbf{p}|\mp
i\epsilon $, in order to avoid these poles in Wick rotation the integral
limit of $p_{0}$ should be $[-i\infty ,i\infty ]$ which corresponds to 
$[-\infty ,\infty ]$ for the $p_{4}$ integral. We have also used 
\begin{equation}
\partial _{\sigma }^{p_{E}}(p_{E}^{\sigma }/p_{E}^{4})=2\pi ^{2}\delta
^{(4)}(p_{E}^{\sigma }).
\end{equation}
Although for $p_{E}\neq 0$, we have $\partial _{\sigma
}^{p_{E}}(p_{E}^{\sigma }/p_{E}^{4})=0$, but the integral is non-vanishing,
since 
\begin{equation}
\int d^{4}p_{E}\partial _{\sigma }^{p_{E}}(p_{E}^{\sigma }/p_{E}^{4})=\oint
dS_{3,\sigma }p_{E}^{\sigma }/p_{E}^{4}=2\pi ^{2}.
\end{equation}
Note that the n-volume of n-sphere or the hyper-surface area of (n+1)-ball
with radius $R$ is given by $S_{n}=[2\pi ^{(n+1)/2}/\Gamma ((n+1)/2)]R^{n}$.
So we see that $\delta (p^{2})b^{\sigma }$ plays the role of the Berry
curvature of a 4-dimensional monopole in Euclidean momentum space, where the
singular point of the monopole is located at $p_{E}=0$. This is related to
the 3-dimensional case by simply imposing the on-shell condition 
$\int dp_{0}\delta (p^{2})b^{\sigma }=\left( 0,\bm{\Omega}/2\right) $.

In fact, Eq.(\ref{wigner}) has encoded the Berry phase already. When a plus
helicity fermion moves under a weak external electromagnetic field, its
momentum changes adiabatically from $k$ to $k^{\prime }$. The Wigner
function could develope a nontrivial Berry phase related to the matrix
element $\left\langle k+\delta k\left\vert \text{tr}\hat{W}\gamma
^{0}\right\vert k\right\rangle \propto u^{\dagger }(k^{\prime })u(k)\simeq
e^{i\delta k\cdot a}$, where we have $u^{\dagger }(k)u(k)=1$\ and $a^{\alpha
}\equiv iu^{\dagger }(k)\partial _{k}^{\alpha }u(k)=\left( a^{0},\mathbf{a}
\right) $. The curvature $\varpi ^{\alpha \beta }\equiv $\ $\partial
_{k}^{\alpha }a^{\beta }-\partial _{k}^{\beta }a^{\alpha }$\ yields a Berry
magnetic field $\varpi ^{ij}=\epsilon ^{ijk}\Omega ^{k}$\ but no Berry
electric field ($\varpi ^{0i}=0$). Thus, the 3-dimensional $\mathbf{a}$\ and $\bm{\Omega}$
are naturally embedded in the 4-dimensional result. 

In summary, we have shown that the Berry curvature and a 4-dimensional
monopole in Euclidean momentum space emerge in a new chiral kinetic equation
with manifest Lorentz covariance. The chiral anomaly can be interpreted as
the flux of this 4-dimensional monopole. There are vorticity terms in this
chiral kinetic equation which are necessary for the presence of the chiral
vortical effect. The 3-dimensional chiral kinetic equation can be obtained
from the Lorentz covariant one by integration over the zero-th component of
the 4-momentum. It contains vorticity terms in addition to what is
previously derived in the Hamiltonian approach. The phase space continuity
equation has an anomalous source term proportional to the product of
electric and magnetic fields. Our approach to the chiral kinetic equation is
quite general and valid for relativistic fermionic systems.

Note added: During the completion of this work, we learned that Son and
Yamamoto were also working on the similar topic \cite{Son:2012zy}.

\textit{Acknowledgment.} This work is supported by the NSFC under grant No.
11125524, 1221504 and 11205150, and by the U.S. DOE under Contract No.
DE-AC02-05CH11231 and within the framework of the JET Collaboration. JWC and
SP are supported in part by the NSC, NTU-CTS, and the NTU-CASTS of R.O.C.


\begin{thebibliography}{99}
\bibitem{Berry:1984} M.~V.~Berry, Proc.\ Roy.\ Soc.\ Lond.\ \textbf{A392},
45 (1984).

\bibitem{Xiao:2010} D.~Xiao, M.-C.~Chang, and Q.~Niu, Rev.\ Mod.\ Phys.\ 
\textbf{82}, 1959 (2010).


\bibitem{Son:2012wh} D.~T.~Son and N.~Yamamoto, 
arXiv:1203.2697 [cond-mat.mes-hall]. 


\bibitem{Stephanov:2012ki} M.~A.~Stephanov and Y.~Yin, 
arXiv:1207.0747 [hep-th]. 

\bibitem{Wong:2011} C.~H.~Wong and Y.~Tserkovnyak, Phys.\ Rev.\ B \textbf{84}%
, 115209 (2011).


\bibitem{Kharzeev:2007jp} D.~E.~Kharzeev, L.~D.~McLerran and H.~J.~Warringa, 
Nucl.\ Phys.\ A \textbf{803}, 227 (2008). 


\bibitem{Fukushima:2008xe} K.~Fukushima, D.~E.~Kharzeev and H.~J.~Warringa, 
Phys.\ Rev.\ D \textbf{78}, 074033 (2008). 


\bibitem{Kharzeev:2010gr} D.~E.~Kharzeev and D.~T.~Son, 
Phys.\ Rev.\ Lett.\ \textbf{106}, 062301 (2011). 


\bibitem{Erdmenger:2008rm} J.~Erdmenger, M.~Haack, M.~Kaminski and A.~Yarom, 
JHEP \textbf{0901}, 055 (2009). 


\bibitem{Banerjee:2008th} N.~Banerjee, J.~Bhattacharya, S.~Bhattacharyya,
S.~Dutta, R.~Loganayagam and P.~Surowka, 
JHEP \textbf{1101}, 094 (2011). 


\bibitem{Torabian:2009qk} M.~Torabian and H.~U.~Yee, 
JHEP \textbf{0908}, 020 (2009). 


\bibitem{Rebhan:2009vc} A.~Rebhan, A.~Schmitt and S.~A.~Stricker, 
JHEP \textbf{1001}, 026 (2010). 


\bibitem{Kalaydzhyan:2011vx} T.~Kalaydzhyan and I.~Kirsch, 
Phys.\ Rev.\ Lett.\ \textbf{106}, 211601 (2011). 


\bibitem{Son:2009tf} D.~T.~Son and P.~Surowka, 
Phys.\ Rev.\ Lett.\ \textbf{103}, 191601 (2009). 


\bibitem{Pu:2010as} S.~Pu, J.~H.~Gao and Q.~Wang, 
Phys.\ Rev.\ D \textbf{83}, 094017 (2011). 


\bibitem{Sadofyev:2010pr} A.~V.~Sadofyev and M.~V.~Isachenkov, 
Phys.\ Lett.\ B \textbf{697}, 404 (2011). 


\bibitem{Lin:2011mr} S.~Lin, 
Nucl.\ Phys.\ A \textbf{873}, 28 (2012). 


\bibitem{Kharzeev:2011ds} D.~E.~Kharzeev and H.~-U.~Yee, 
Phys.\ Rev.\ D \textbf{84}, 045025 (2011). 


\bibitem{Metlitski:2005pr} M.~A.~Metlitski and A.~R.~Zhitnitsky, 
Phys.\ Rev.\ D \textbf{72}, 045011 (2005). 


\bibitem{Newman:2005as} G.~M.~Newman and D.~T.~Son, 
Phys.\ Rev.\ D \textbf{73}, 045006 (2006). 


\bibitem{Charbonneau:2009ax} J.~Charbonneau and A.~Zhitnitsky, 
JCAP \textbf{1008}, 010 (2010). 


\bibitem{Lublinsky:2009wr} M.~Lublinsky and I.~Zahed, 
Phys.\ Lett.\ B \textbf{684}, 119 (2010). 


\bibitem{Asakawa:2010bu} M.~Asakawa, A.~Majumder and B.~Muller, 
Phys.\ Rev.\ C \textbf{81}, 064912 (2010). 


\bibitem{Landsteiner:2011cp} K.~Landsteiner, E.~Megias and F.~Pena-Benitez, 
Phys.\ Rev.\ Lett.\ \textbf{107}, 021601 (2011). 


\bibitem{Hou:2012xg} D.~-F.~Hou, H.~Liu and H.~-c.~Ren, 
Phys.\ Rev.\ D \textbf{86}, 121703 (2012). 


\bibitem{Gao:2012ix} J.~-H.~Gao, Z.~-T.~Liang, S.~Pu, Q.~Wang and
X.~-N.~Wang, 
Phys.\ Rev.\ Lett.\ \textbf{109}, 232301 (2012). 


\bibitem{Xiao:2005qw} D.~Xiao, J.~-r.~Shi and Q.~Niu, 
Phys.\ Rev.\ Lett.\ \textbf{95}, 137204 (2005) [cond-mat/0502340].


\bibitem{Vasak:1987um} D.~Vasak, M.~Gyulassy and H.~T.~Elze, 
Annals Phys.\ \textbf{173}, 462 (1987). 


\bibitem{Elze:1986qd} H.~T.~Elze, M.~Gyulassy and D.~Vasak, 
Nucl.\ Phys.\ B \textbf{276}, 706 (1986). 


\bibitem{Elze:1989un} H.~T.~Elze and U.~W.~Heinz, 
Phys.\ Rept.\ \textbf{183}, 81 (1989). 


\bibitem{Son:2012zy} D.~T.~Son and N.~Yamamoto, 
arXiv:1210.8158 [hep-th]. 
\end{thebibliography}
\end{document}